\documentclass[twocolumn,showpacs,preprintnumbers,amsmath,amssymb]{revtex4}
\usepackage{pifont}
\usepackage{tipa}
\usepackage{graphicx}
\usepackage{dcolumn}
\usepackage{bm}
\usepackage{multirow}

\begin{document}

\title{Scalar glueball in a soft-wall model of AdS/QCD}

\author{Xue-Feng Li and Ailin Zhang\footnote{corresponding author:
zhangal@staff.shu.edu.cn}} \affiliation{Department of Physics,
Shanghai University, Shanghai 200444, China}

\begin{abstract}
 The scalar glueball is investigated in a soft-wall model of AdS/QCD. Constraints of the mass of the scalar glueball are given through an analysis of a relation between the bulk mass and the anomalous dimension. The mass of the ground scalar glueball is located at $0.96^{+0.04}_{-0.07}~\rm{GeV}<m_G<1.36^{+0.05}_{-0.10}~\rm{GeV}$. In terms of a background dilaton field $\Phi(z)=cz^2$, the two-point correlation function for the scalar gluon operator is obtained. The two-point correlation function at $\Delta=4$ gives a different behavior compared with the one in QCD.
\end{abstract}


\maketitle

\section{Introduction}

The self-interaction among gluons is a distinctive feature
in QCD. Bound gluon states, glueballs, may exist in QCD, and were first mentioned by Fritzsch {\it et al.}\cite{fritzsch}
Though the properties of glueballs can not be obtained analytically through QCD,
they are predicted in many models based on QCD or in lattice QCD. Therefore, the discovery of a glueball will be a direct test of QCD
theory. People has searched for the glueball for a long time. So far, there is no conclusive evidence of glueball except for several observed candidates.

No matter whether glueball exists or not, it is important to understand their spectroscopy. So far, there are not so many work concentrated on the productions and decays of glueballs, but there are many work on the masses of glueballs. Related references can be found in some reviews and therein~\cite{review,review1,review2}. It is believed that the $J^{PC}=0^{++}$ scalar glueball has the lowest mass. The first assumption of the mass $M$ of the scalar glueball is by Novikov {\it et al.,} where $M$ was assumed around $700$ MeV\cite{novikov}. QCD sum rules predict the mass of this kind of glueball around $1.5$ GeV~\cite{sum,sum1,sum2}, and lattice theory gives the mass around $1.7$ GeV~\cite{lattice}.

Inspired by the gravity/gauge, or anti-de Sitter/conformal field theory (AdS/CFT) correspondence~\cite{AdS}, QCD is in one side modeled by a deformation of the super Yang-Mills theory (so called top-down~\cite{top} approach). On the other side, a five dimensional holographic QCD is constructed from the same correspondence (so called bottom-up~\cite{bottom} approach). In the framework of holographic QCD,
some unsolvable problems in QCD are solvable. In particular, the hadrons of QCD correspond to the normalizable modes of the five dimensional (5d) fields, and therefore many hadronic properties are predictive.

In AdS/QCD, two backgrounds have been introduced: the hard wall approach (an IR brane cutoff)~\cite{hard wall} or the soft wall approach (a soft wall cutoff)~\cite{soft wall}. For the soft wall cutoff, the dilaton soft wall and the metric soft wall are usually employed. The dilaton soft wall model with smooth cutoff seems suitable for the scalar glueball, which reproduces the expected Regge trajectory.

The mass spectrum of the scalar glueball has been computed in the holographic QCD. In Ref.~\cite{Colangelo}, the mass square of the scalar glueball is predicted twice of the mass square of the $\rho$ meson: $m^2_G=2m^2_\rho$. In Ref.~\cite{forkel}, the scalar glueball mass is predicted at $1.34$ GeV or $1.80$ GeV corresponding to Neumann or Dirichlet boundary conditions on the IR brane, respectively. Some theoretical predictions to the mass of scalar glueball are listed in Table. 1. Obviously, the mass prediction which is waiting for an examination by experiment may be largely different in different models.
\begin{center}
\begin{table*}
\begin{tabular}{ccccccccc}
\hline
\hline
Literature & Ref.~(\cite{novikov}) & Ref.~(\cite{sum}) & Ref.~(\cite{sum1}) & Ref.~(\cite{tom}) & Ref.~(\cite{sum2}) & Ref.~(\cite{lattice})  & Ref.~(\cite{vento}) & Ref.~(\cite{forkel}) \\
Mass (GeV) & 0.7    & $1.5\pm 0.2$    & 1.7 & 1.4 and 1.0-1.25   & $1.25\pm 0.2$    & 1.71     & 0.65-0.75    & 1.34 or 1.80      \\
\hline\hline
\end{tabular}
\caption{Mass predictions of the scalar glueball in some literature} \label{table-1}
\end{table*}
\end{center}

The two-point correlation function of the scalar operator has been analyzed in Refs.~\cite{forkel,glueball}, where the behavior of the correlation function is given and compared with the result of the operator product expansion (OPE) in QCD.

In this paper, we focus on the properties of the scalar glueball in
the framework of a dilaton soft-wall. We predict the possible
mass of the scalar glueball in section 2. In section 3, we give the result of the two-point correlation
function for the scalar glueball with $\Phi(z)=cz^2$ . The conclusions are included in the final section.

\section{Mass of scalar glueball in dilaton soft wall model}

According to AdS/CFT correspondence, the correspondence between QCD local operators and fields in the $AdS_5$ bulk space can be constructed, and there exists following relation~\cite{AdS,Colangelo}:
$$m^2_5=(\triangle-P)(\bigtriangleup+P-4)$$
where $m_5$ is the AdS mass of the dual field in the bulk, $\Delta$ is the conformal dimension of a (p-form) operator.

It is noticed that the full conformal dimension $\Delta$ could be divided into a classical
dimension $\Delta_{class}$ and an anomalous dimension $\gamma(\mu)$~\cite{Boschi}: $$\Delta=\Delta
_{class}+\gamma(\mu).$$

The operator $tr\{G_{\mu\nu}G^{\mu\nu}\}$ defined on the boundary
spacetime is related to the scalar glueball. Following Gubser's proposal~\cite{Gubser}, the full dimension of operator $tr
G^2$(in correspondence to the scalar glueball) is
$$\Delta
_{G^2}=4+\beta '(\alpha )-\frac{2 \beta (\alpha )}{\alpha },$$
where the prime is denoted to the derivative with respect to $\alpha$.
This expression for scalar glueballs emerges from the trace anomaly
of QCD energy momentum tensor $T_{\mu }^{\mu }$.

The $5d$ massive scalar bulk field $X$ in the soft-wall model of AdS/QCD can be described by an action~\cite{Boschi}
\begin{eqnarray}\label{eq1}
S_{5D}=\int  d^5x\sqrt{-g}e^{-\Phi (z)}\left[g^{mn}\partial
_mX\partial _nX+m_5^2X^2\right],
\end{eqnarray}
where the background dilaton field $\Phi(z)=cz^2$ and $g$ is the determinant of the metric tensor in $AdS_5$ space.

The metric is defined as follows
\begin{eqnarray*}
ds^2\equiv
g_{mn}dx^mdx^n=\left(\frac{R}{z}\right)^2\left(\eta
_{\mu \nu }dx^{\mu }dx^{\nu }+dz^2\right),
\end{eqnarray*}
where $\eta_{\mu,\nu}$=(-,+,+,+) is the Minkowski metric. In the following,
the AdS radius R is assumed to be unit.

From the action Eq.~(\ref{eq1}), the $1d$ sch\"{o}dinger-type equation of motion is obtained~\cite{Boschi}
\begin{eqnarray}\label{eq2}
 -\tilde{Y}{''}[z]+\left(c^2z^2+\frac{15}{4 z^2}+2c+\frac{m_5{}^2}{ z^2}\right)\tilde{Y}[z]=-q^2 \tilde{Y}[z],
\end{eqnarray}
where the Bogoliubov transformation $\tilde{X}(q,z)=e^{\frac{\left(cz^2+3\ln z\right)}{2}} \tilde{Y}[z]$ is performed. Here $\tilde{X}$ is the $4d$ Fourier transform of the field X. Through this $1d$ sch\"{o}dinger-type equation of motion, the spectrum is given $m_G^2=4cn+4c+2c\sqrt{4+m_5^2}$~\cite{Colangelo}.

In the classic case,
$\Delta_{class}=4$, ${m_5}^2=0$. Therefore the mass of the scalar
glueball is $m^2_{G}=4c (n+2)$~\cite{Boschi}, where $n$ is
identified with the radial quantum number. When the anomalous
dimension of operator $tr\{G_{\mu\nu}G^{\mu\nu}\}$ is taken into account, the conformal dimension is shifted.
Such an effect was explored in Ref.~\cite{Boschi}, where the mass
dependence of the Beta function in holographic QCD was also discussed.

The explicit expressions of the anomalous dimension and hence the bulk
mass $m_{5}(z)$ in holographic QCD are usually difficult to obtain. When the anomalous dimension is taken into account, the bulk mass is shifted to
\begin{eqnarray}
 m_5^2(z)=\gamma (z)(\gamma (z)+4).
\end{eqnarray}

Therefore, the corrected potential with this $\gamma$ correction in the $1d$ sch\"{o}dinger-type equation becomes
\begin{eqnarray}
\Delta V(z)=\frac{\gamma (z)(\gamma (z)+4)}{z^2}.
\end{eqnarray}
It is straightforward to obtain an inequity
$$\Delta V(z)\geq \frac{-4}{z^2},$$
where the inequity is saturated for the constant dimension $\Delta=2$
or $\gamma$=-2. Similar analysis was performed to tetraquarks in Ref.~\cite{tetraquarks}.

Obviously, in such a case, the Breitenlohner-Freedman relation~\cite{BF}
\begin{eqnarray}
m_5^2>-\frac{(d-2P)^2}{4}
\end{eqnarray}
holds also, where the scalar and the vector correspond to $P=0$ and $P=1$, respectively~\cite{Bianchi}. As well known, to keep the ground state wave-function integrable, the
Breitnlohner-Freedman (BF) bound for the bulk mass ($\Delta
=\frac{d}{2}+\sqrt{\frac{(d-2P)^2}{4}+m_5^2}$~\cite{dimension}) must be satisfied.

Accordingly, the lowest limit for the ground ($n=0$) scalar
glueball mass is achieved
\begin{eqnarray}
m_G^2{}_{,\Delta =2,n=0}=4c.
\end{eqnarray}

On the other hand, the highest limit for the ground scalar glueball mass is
$$m_G{}_{,\Delta =4,n=0}=2\sqrt{2c}.$$
Once $c$ is fixed by the mass spectrum of the vector $\rho$ mesons in AdS/QCD with $c=0.2325$ GeV$^2=(0.482$~GeV)$^2$\cite{Boschi}, the possible mass of the ground scalar glueball becomes
$$0.96~\rm{GeV}<m_G<1.36~\rm{GeV}.$$

Of course, there is an uncertainty for the parameter $c$, which will result in an uncertainty to the mass. From a fitting result for the Regge trajectories of vector and axial-vector mesons $\rho$ and $a$, $c$ is found around $0.20-0.25$~\cite{yan}. When $c=0.2325$ GeV$^2=(0.482$~GeV)$^2$ is employed as the central value and the uncertainty of $c$ is taken into account, the possible mass of the ground scalar glueball is
$$0.96^{+0.04}_{-0.07}~\rm{GeV}<m_G<1.36^{+0.05}_{-0.10}~\rm{GeV}.$$

\section{Two-point correlation function for scalar glueball}

In the conjectured AdS/CFT correspondence, $4d$ field is
supposed to locate at the boundary of $AdS_5$. The bulk field coupling to an operator on the boundary
could be detected through the interaction \^{O}$X _0$ in the lagrangian. One can
compute the correlation function on the boundary according to the
Gubser-Klebanov-Polyakov-Witten relation (GKPW)~\cite{AdS}
\begin{eqnarray}
\left\langle e^{i\int d^4x X_0\hat{O(x)}} \right\rangle
{}_{CFT}=e^{i\int S_{5D} [X]}
\end{eqnarray}
where $S_{5D}$ is defined in Eq.(\ref{eq1}), and the index zero on the scalar field
denotes the source on boundary. The scalar field X(x,z)
is coupled to the lowest dimension operator
\^{O}=$\beta(\alpha)Tr(G^2)$ for the scalar glueball.

Following the process in Refs.~\cite{forkel,glueball}, the two-point correlation
function in AdS is obtained from the action as follows
\begin{eqnarray}
&&\left.\Pi _{AdS}\left(q^2\right)=2 \frac{ e^{-c z^2}}{z^3}
\tilde{K}(q,z)\partial _z\tilde{K}(q,z)\right|{}_{z\to
0}, \nonumber\\
&&+2m_5^2\frac{ e^{-c z^2}}{z^5}
\left. \left(\tilde{K}(q,z)\right)^2 \right|_{z\to 0},
\end{eqnarray}
where $\tilde {K}(q,z)$ is the momentum representation of the bulk-to-boundary propagator $K(x,z)$ which satisfies the same equation Eq.~(\ref{eq2}) with the bulk field $X$. Here
\begin{eqnarray}{rr}
&& X(x,z)=\int
d^4x'K\left(x-x',z\right)X_0\left(x'\right)
\end{eqnarray}
with ${X}_0(q)$=$ X(x,z\rightarrow0)$.

At $\Delta =4$ (${m_5}^2=0$), we have
\begin{eqnarray}\label{correlation}
&&\Pi  _{AdS,\Delta =4}\left(q^2\right)=-q^2 \upsilon ^2-\frac{1}{4} q^2 \left(4 c+q^2\right)\left(-2+\gamma _E\right)\nonumber\\&&-\frac{1}{4} q^2 \left(4 c+q^2\right)HN\left[1+\frac{q^2}{4 c}\right]-\nonumber\\&&\frac{1}{4} q^2 \left(4 c+q^2\right)\left(Log\left[c\right]+2 Log\left[\frac{1}{\upsilon }\right]\right),
\end{eqnarray}
where z has been replaced by the renormalization scale $\upsilon ^{-1}$, and HN indicates the Harmonic number.

In the limit $q^2\to \infty$, the two-point correlation function in the short-distance regime transfers into
\begin{eqnarray}
 &&\Pi _{AdS,\Delta =4}\left(q^2\right)=\frac{1}{4} \left(2-2 \gamma _E+Log[4]- Log\left[\frac{q^2}{\upsilon ^2}\right]\right)q^4+\nonumber\\&&\frac{ cq^2}{2}-\upsilon ^2q^2+c q^2 Log[4]-2c q^2\gamma _E+c q^2 Log\left[\frac{\upsilon ^2}{q^2}\right]\nonumber\\&&-\frac{5 c^2}{3}+\frac{4 c^3}{3 q^2}-\frac{8 c^4}{15 q^4}+O\left[\frac{1}{q^6}\right].
\end{eqnarray}
Similar Result has been obtained in Ref.~(\cite{glueball}), where the authors discussed the important difference between the AdS expression and the one in QCD in detail. The behavior of the two-point correlation function is presented in figure 1, where the renormalization scale is chosen with $\nu$=1 GeV and the scale parameter is chosen with $c=0.233$ GeV$^2$. In the figure, the poles at the region $q^2<0$ give the mass spectrum $m^2_{G}=4c (n+2)$~\cite{Colangelo,glueball} with $n$ indicating the radial quantum number. The behavior of the two-point correlation function in QCD for the scalar glueball is also presented in the figure according to Eq. (2.9) in Ref.~(\cite{glueball}), where the parameters are those from Ref.~(\cite{paver}).

In the limit $c\to 0$, namely in the pure anti-de Sitter bulk space, we have a massless boundary theory.
In the case of a massive scalar bulk field background with a dilaton field, there appears a
term of $q^2$ involving dimension two condensate. The detailed discussions
about this term have been made in Refs.~(\cite{forkel,glueball,fangzuo}). As mentioned by
Colangelo~\cite{glueball}, such a term does not appear in the QCD
short-distance expansion, and can not be expressed as vacuum
expectation value of local operators.

\begin{center}
\begin{figure}
\includegraphics[width=8cm,keepaspectratio]{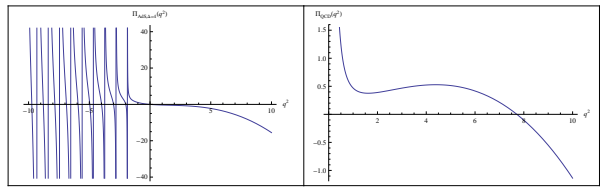}
\caption{Behavior of the two-point correlation functions for the scalar glueball in AdS and QCD.}
\end{figure}
\end{center}

\section{Conclusion}

The scalar glueball is investigated in the framework of a soft-wall
model of AdS/QCD. Through the relation between the bulk mass and the anomalous dimension, we give the constraints on the mass of the scalar glueball. Our results indicate a light scalar glueball. The possible mass of the ground scalar glueball is located at the mass region: $0.96^{+0.04}_{-0.07}~\rm{GeV}<m_G<1.36^{+0.05}_{-0.10}~\rm{GeV}$. Obviously, the result depends on the choice of the parameter $c$ in AdS/QCD. In order to predict the mass of the scalar glueball, the parameter $c$ has to be fixed more precisely.

The dilaton field is chosen as $\Phi(z)=cz^2$, and the two-point correlation function for the scalar glueball is obtained. In particular, the exact expression of the correlation function is presented at $\Delta =4$, which shows a different behavior compared with the one in QCD. Phenomenological results related to the two-point correlation for the scalar glueball deserve dedicated exploration.

Acknowledgment: This work is supported in part by National Natural
Science Foundation of China(11075102) and the Innovation Program of Shanghai Municipal Education Commission under grant No. 13ZZ066.

\vspace{5mm}
\clearpage
\end{document}